\documentclass[prb,aps,twocolumn,floats,superscriptaddress]{revtex4-1}
\include{header}
\usepackage{soul}
\usepackage{xcolor}
\usepackage{multirow}
\usepackage{amsmath}
\usepackage{graphicx}
\usepackage{amssymb}
\usepackage{epsfig}
\usepackage{simplewick}
\usepackage{esint}
\usepackage{ulem}
\usepackage[colorlinks = true]{hyperref}
\newlength{\upit}\upit=0.1truein

\newcommand{\ltappr}{{{\lower4pt\hbox{$<$} } \atop \widetilde{ \ \ \ }}}
\newlength{\bxwidth}\bxwidth=1.5 truein

\newcommand{\tr}{{\hbox{Tr}}}

\newcommand{\dg}{^{\dagger }}

\newcommand{\gtappr}{{{\lower4pt\hbox{$>$} } \atop \widetilde{ \ \ \ }}}

\newcommand{\bk}{{\bf{k}}}

\newcommand{\be}{\begin{equation}}
	\newcommand{\ben}{\begin{equation*}}
		\newcommand{\ee}{\end{equation}}
	\newcommand{\een}{\end{equation*}}

\newcommand{\bmx}{\begin{array}}
	\newcommand{\emx}{\end{array}}
\newcommand{\bean}{\begin{eqnarray*}}
	\newcommand{\eean}{\end{eqnarray*}}

\newcommand{\sgn}[1]{{\rm sign}{#1}}

\newcommand{\mat}[1]{\left(\bmx{cc}#1\emx\right)}

\newcommand\ltdash{\raise-0.7pt\hbox{$\scriptscriptstyle |$}}

\newlength{\figwidth}
\newlength{\shift}
\begin{document}
	\title{Analytic calculation of the vison gap in the Kitaev
	spin liquid
	}
	\author{Aaditya Panigrahi}
	\affiliation{
		Center for Materials Theory, Department of Physics and Astronomy,
		Rutgers University, 136 Frelinghuysen Rd., Piscataway, NJ 08854-8019, USA}
	\author{Piers Coleman}
	\affiliation{
		Center for Materials Theory, Department of Physics and Astronomy,
		Rutgers University, 136 Frelinghuysen Rd., Piscataway, NJ 08854-8019, USA}
	\affiliation{Department of Physics, Royal Holloway, University
		of London, Egham, Surrey TW20 0EX, UK.}
	
	\date{\today}
	\author{Alexei Tsvelik}
\affiliation{Division of Condensed Matter Physics and Materials Science, Brookhaven National Laboratory, Upton, NY 11973-5000, USA}
	
	\pacs{PACS TODO}
	\begin{abstract}
		{Although the ground-state energy of the Kitaev spin
		liquid can be calculated exactly, the associated 
		vison gap energy has to date only been calculated numerically from		finite size diagonalization. Here we show that the
		phase shift for scattering Majorana fermions off a
		single bond-flip can be calculated analytically,
		leading to a closed-form expression for the vison gap
		energy $\Delta =0.2633 J$. 
        Generalizations of our approach
		can be applied to  Kitaev spin liquids on more
		complex lattices  such as the three dimensional
		hyper-octagonal lattice.  
		}
	\end{abstract}
	\maketitle
	
	\section{Introduction}
\begin{figure}[bh]
		\includegraphics[scale=.25]{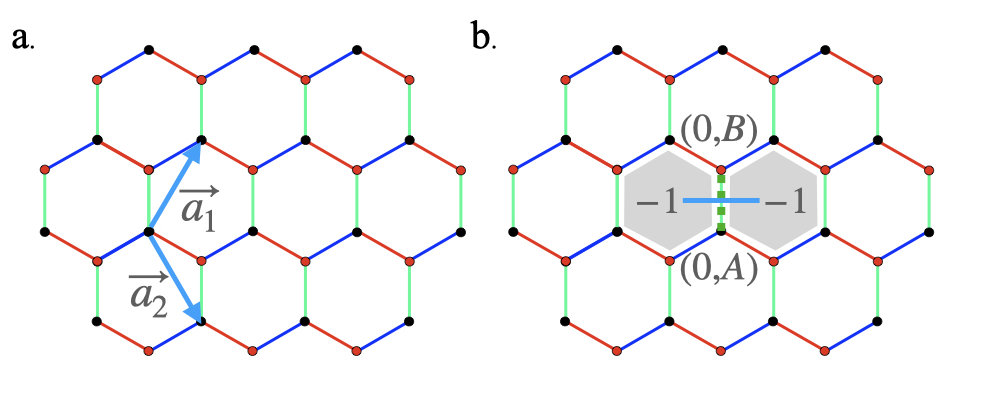}
		\caption{(a) The Kitaev honeycomb lattice
		model, where the Ising spin couplings
			along the x, y and z
			directions are labelled by blue, green and red bonds
			respectively, with primitive lattice vectors
			$\vec{a}_1$ and $\vec{a}_2$. (b) 
A bond-reversal at the origin creates a vison pair, costing an energy $\Delta_v$. The string connecting the
adjacent visons is indicated in light blue.}
		\label{KSLFig}
	\end{figure}

Kitaev spin liquids (KSL) are a class
of exactly solvable quantum spin liquid that exhibit spin
fractionalization, anyonic excitations and long-range
entanglement\cite{KSL1,hermanns_physics_2018,obrien_classification_2016,trebst_kitaev_2017,takagi_concept_2019}. The fractionalization of spins into Majorana fermions is accompanied by the formation of emergent $\mathbb{Z}_2$
gauge fields, giving rise to 
 $\mathbb{Z}_2$ vortex excitations or ``visons''. These excitations are gapped, and the energy cost associated with creating two visons on adjacent plaquettes is called the vison gap $\Delta_v$ (Fig[\ref{KSLFig}]).  
Proposals
for the practical realization of Kitaev spin liquids in quantum
materials, including 
$\alpha$-RuCl$_3$
\cite{banerjee_proximate_2016,do_majorana_2017,jackeli_mott_2009,kasahara_majorana_2018,liu_kitaev_2020,takagi_concept_2019,winter_models_2017,wolter_field-induced_2017,wulferding_magnon_2020,yamada_andersonkitaev_2020}
and Iridates
\cite{chaloupka_kitaev-heisenberg_2010,kitagawa_spinorbital-entangled_2018,winter_models_2017}
have renewed interest in the thermodynamics of Kitaev spin
liquid
\cite{eschmann_thermodynamics_2019,feng_further_2020,joy_dynamics_2022,kato_chiral_2017,li_universal_2020,nasu_thermal_2015,motome_hunting_2020,udagawa_spectroscopy_2019}. The 
extension of these ideas to Yao-Lee spin liquid
\cite{yao_exact_2007,yao_fermionic_2011} and its application to Kondo
models, \cite{coleman_solvable_2022,tsvelik_order_2022} motivate the
development of an analytical approach to calculate the vison gap
$\Delta_v$. 

The vison gap in KSLs
has to date, been determined
by numerical diagonalization of finite size systems
\cite{obrien_classification_2016,KSL1}. Here we present a Green's
function approach for the analytical computation of the vison gap
$\Delta_v$ from the scattering phase shift associated with a
a $\mathbb{Z}_2$ bond-flip. 
 Our work builds on theoretical developments in the field of Kitaev
 spin liquids which relate to the interplay between Majorana
 fermions and
 visons\cite{KSL1,baskaran_exact_2007,knolle_dynamics_2014,lahtinen_interacting_2011,lahtinen_perturbed_2014,lahtinen_spectrum_2008,lahtinen_topological_2012,theveniaut_bound_2017,KSL1,joy_dynamics_2022,kao_vacancy-induced_2021,knolle_dynamics_2014,lahtinen_interacting_2011,lahtinen_perturbed_2014,lahtinen_spectrum_2008,lahtinen_topological_2012,theveniaut_bound_2017,zhang_vison_2019}.
Using exact calculations, we find the
vison gap energy of $\Delta_v=0.263313(6) J$ for the Kitaev spin liquid on
honeycomb lattice in the gapless phase, extending the accuracy of
previous calculations
\cite{KSL1,obrien_classification_2016}. 
Our calculations reveal the formation of Majorana resonances in the
density of states which accompany the 
formation of two adjacent visons. {Our approach can be
simply generalized to more complex lattices and 
are immediately generalizable to Yao-Lee spin liquids. }

	\section{Vison Gap in the Kitaev Honeycomb Model}
\begin{figure}[h]
	\includegraphics[width=\columnwidth]{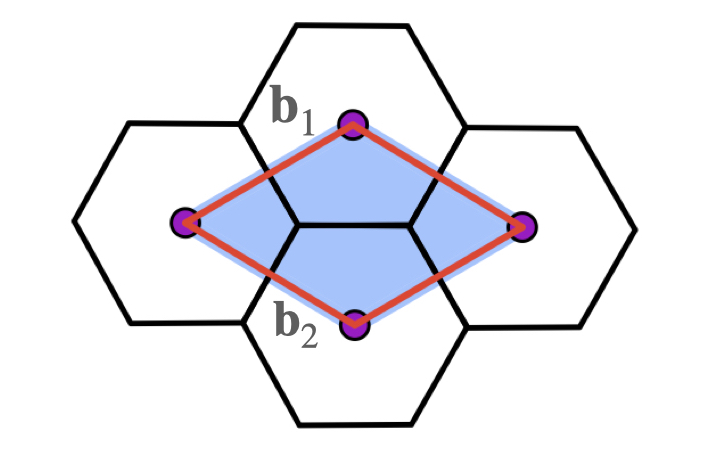}
	\caption{Rearranged 
first Brillouin zone (${\bf BZ}$) constructed in the reciprocal lattice vector basis spanned by ${\bf b_1}$ and $\bf{b_2}$.}\label{Bzone}
\end{figure}

	The Kitaev honeycomb lattice model \cite{KSL1} is described by the Hamiltonian
	\begin{equation}
		H_{K}= \sum_{<ij>} J_{\alpha_{ij}}\sigma^{\alpha_{ij}}_{i}\sigma^{\alpha_{ij}}_{j},
		\label{KSLHam}
	\end{equation}
	where the Heisenberg spins 
 $\vec{\sigma}_{i}= (\sigma^{x}_{i}, \sigma^{y}_{i},\sigma^{z}_{i})$ 
at site $i$ interact with their
 nearest neigbors via an Ising coupling between the 
 $\alpha_{ij}=x,y,z $ spin components, 
along the coresponding bond directions $\langle ij\rangle $,  with strength
	$J_{\alpha_{ij}}$, 
as shown in Fig[\ref{KSLFig}]. An exact
	solution of Kitaev Model \cite{KSL1} is found by representing
	the spins as products of Majorana fermions,
	$\sigma^{\alpha}_j=2 c_j b^{\alpha}_j$ which satisfy
	canonical anticommutation algebras, $\{c_{i},b^{\alpha }_{j}\}=0$,
	$\{b^{\alpha}_i,b^{\beta}_j\}=\delta_{ij}\delta^{\alpha,\beta}$, 
	(taking the convention that $c_{j}^{2}= (
	b_{j}^{\alpha })^{2}=1/2$). The system is projected into the
	physical subspace by selecting $ \mathcal{D}_j\equiv -4i c_j
	b^x_j b^y_j b^z_j =1$ at each site, allowing the Hamiltonian
	(\ref{KSLHam}) to be rewritten as $\mathbb{Z}_{2}$ gauge theory 
	\begin{equation}
		H_{KSL}=2\sum_{<ij>} J_{\alpha_{ij}}\hat{u}_{ij}(i c_i c_j),
		\label{MajoranaKSLHam}
	\end{equation}
where the gauge fields $
\hat{u}_{ij}=2ib^{\alpha_{ij}}_ib^{\alpha_{ij}}_j=\pm 1$ on bond $ij$
commute with the Hamiltonian, 
$[\hat{u}_{ij},H_K]=0$.
The plaquette
operators $\mathcal{W}_p$ 
\begin{equation} \mathcal{W}_p=\prod_{<i,j>\in p} u_{ij}
\quad(i\in A, j\in B).  \label{Plaquette} \end{equation}
formed from the 
product of gauge fields
$\hat{u}_{ij}$ around the hexagonal loop $p$ (
plaquette), are gauge invariant
and also commute with the Hamiltonian
$[\mathcal{W}_p,H_K]=0$ and constraint operators
$[\mathcal{W}_p,\mathcal{D}_j]=0$, giving rise to a set of 
static constants of
motion which take values $\mathcal{W}_p=\pm 1$. Each eigenstate is
characterized by the configurations of $\{\mathcal{W}_p\}$; Lieb's
theorem \cite{lieb_flux_1994} specifies that the ground state
configuration is flux-free, i.e. $\mathcal{W}_p=1$ for all hexagons
$p$. In what follows we will choose the gauge $\hat{u}_{ij}=1$ when
$i\in A$ and $ j\in B$ sublattice, assigning 
\begin{equation}\label{}
H_{0}=
H_{KSL}[u_{ij}\rightarrow 1].
\end{equation}
Rewriting $H_{0}$ in momentum space, we obtain 
\begin{equation}\label{}
H_{0 }= \frac{1}{2}\sum_{{\bf k}\in {\bf BZ}} {\psi }^{\dagger}_{{\bf
k}}(\vec{\gamma}_{{\bf k}}\cdot \vec{\tau}) \psi _{{\bf k}},
\end{equation}
where  
\begin{equation}\label{}
{\psi _{\bf k}}=
 \frac{1}{\sqrt{N_{c}}}\sum_{j}
\mat{	c_{j,A}\cr c_{j,B}}
e^{-i \bk \cdot {\bf R}_{j}}
\end{equation}
 describes a Majorana in momentum space, where $N_{c}$ is the number of
unit cells and ${\bf R}_{j}$ is the location of the unit cell and 
$\vec{\gamma}_{{\bf k}}=( Re({\gamma_{\bf
	k}}),- Im(\gamma_{\bf k}))$ is expressed in terms of the form
	factor 
	\begin{align}
		\begin{split}
			\gamma_{\bf k}&=2i (J_{z} + J_{x}e^{i k_1}+J_{y}e^{i k_2 }),
			\\{\bf k}&=\frac{k_1}{2\pi}{\bf b_1}+\frac{k_2}{2\pi}{\bf b_2}, \quad k_1,k_2 \in [0,2\pi].
			\label{kBLZ}
		\end{split}
	\end{align}  
Here we have employed a reciprocal lattice basis ${\bf b_1,b_2}$ to
span the momentum ${\bf k}\in {\bf BZ}$, which transforms to rhombus
shaped Brillouin zone in the reciprocal lattice (see Fig[\ref{Bzone}]). 
The Majorana excitation spectrum of the 
Kitaev spin liquid is given by the eigenvalues of $H_{0}$, 
	$\epsilon_{\bf k}=\pm \vert \gamma_{\bf k}\vert$. 

We create two adjacent
	visons by flipping the gauge field in
	the unit cell at origin to $\hat{u}_{(0,A)(0,B)}=-1$ as shown in Fig [\ref{KSLFig}],  resulting in the following Hamiltonian:
	\begin{equation}		\label{ksl_gs}
		H_{KSL+2v}=H_{0} +\hat V, 
	\end{equation}
where
\begin{equation}\label{}
\hat V = -4 J_{z} (i c_{0,A}c_{0,B})
\end{equation}
acts as a scattering term for Majoranas in the
bulk.  In this way, the vison gap calculation is formulated as
a scattering problem. 

For this case, the Hamiltonian is given by 
	\begin{equation}\label{SHam}
		\small	H_{KSL+2v}=
\frac{1}{2}\sum_{{\bf k}\in {\bf BZ}} {\psi }^{\dagger}_{{\bf
k}}(\vec{\gamma}_{{\bf k}}\cdot \vec{\tau}) \psi _{{\bf k}}
+\frac{1}{2}{\bf c}^T_{0} (V\tau_2 ){\bf c}_{0},
	\end{equation}
\begin{equation}\label{}
{\bf c}_0=\mat{	c_{0,A}\cr c_{0,B}}=
\frac{1}{\sqrt{N_c}}\sum_{{\bf k}\in {\bf
	BZ}}{\psi _{\bk }}
\end{equation}
describes a Majorana fermion at the origin and $V=4J_{z}$. 

We now set up the scattering problem in terms of Green's functions.
The Green's function of the unscattered Majoranas is
$G_0= G_0(i\omega_{n},{\bf k})\delta_{{\bf k},{\bf k'}}$, where 
\begin{equation}\label{}
G_{0} (i\omega_{n},\bk  )= [i\omega_{n}-\vec{\gamma}_{\bk }\cdot \vec{\tau }]^{-1}.\end{equation}
In the presence of the bond-flip at the origin, the 
Green's function of the scattered Majoranas is given by 
$G= (G_{0}^{-1}-\hat{V})^{-1}$,  where 
$\hat{V}_{{\bf k},{\bf k'}}=(V\tau_2)/N_c$ is the scattering matrix.
The total free energy of the non-interacting 
ground-state in the presence of the
scattering is given by the standard formula
	\begin{equation}
		\beta F=-\frac{1}{2}\tr[\ln(-G^{-1})]=-\frac{1}{2}\tr\ln[-G^{-1}_0+\hat{V}],
		\label{FreeEnergy}
	\end{equation}
where  ${\rm Tr}  $  denotes the full trace over Matsubara frequencies, 
 momenta and  sublattice degrees of freedom. 
The change in free energy is then given by 
\begin{eqnarray}
\beta \Delta F=-\frac{1}{2 }\tr[\ln(1-\hat{V}G_0)]=
 \frac{1}{2 }\sum_{r=1}^{\infty }\frac{1}{r}{\rm Tr}\bigl [(\hat
VG_{0})^{r}\bigr ].
\end{eqnarray}
We now carry out the trace over 
the Matsubara frequencies and momenta,
so that 
\begin{equation}\label{}
\Delta F= \frac{1}{2\beta }\sum_{i\omega_{n}}\sum_{r=1}^{\infty }\frac{1}{r}{\rm tr}\biggl[\left(
 \frac{V\tau_{2}}{N_{c}}\sum_{\bk }G_{0} (i\omega_{n},\bk )\right)^{r}\biggr],
\end{equation}
where ${\rm tr}\bigl[\ \  \bigr]$ denotes the residual trace over sublattice degrees of
freedom.
Now, we can 
incorporate the summations over
momentum by introducing 
the local Green's function 
	\begin{equation}
		g(z)=\frac{1}{N_c}\sum_{{\bf k} \in \mathbf{BZ}} G_0(z,{\bf k}),
	\end{equation} 
so that 
\begin{eqnarray}\label{}
\Delta F &=& \frac{1}{2\beta }\sum_{i\omega_{n}}\sum_{r=1}^{\infty }\frac{1}{r}{\rm tr}\left[(
V\tau_{2}g (i\omega_{n}))^{r}\right]\cr
&=& -\frac{1}{2 \beta}\sum_{i\omega_{n}} {\rm tr}[\ln(1-V\tau_{2}g (i\omega_{n}))],
\end{eqnarray}
where we have re-assembled the Taylor series
as a logarithm. 
\begin{figure}[tb]
	\includegraphics[scale=.27]{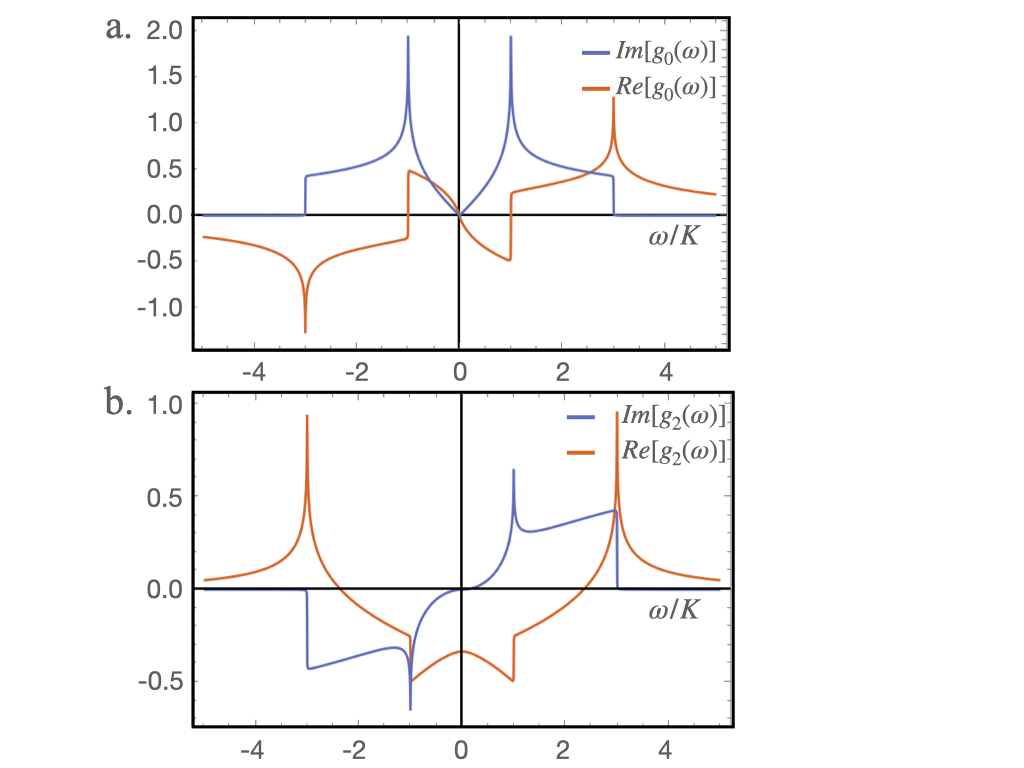}
	\caption{Showing real and imaginary parts of (a)$g_0(\omega-i \delta)$
	and (b) $g_2(\omega-i \delta)$ as defined in equations
	\eqref{GreensFunctionDecompositionX}, \eqref{intG0} and \eqref{intG2}.
}
	\label{GreensFunctionFig}
\end{figure}

We shall illustrate our method for the isotropic case
$J_{x}=J_{y}=J_{z}=J$, setting $K = 2J$ and $V=4J$. In this case, 
$\gamma_{\bf k}=iK (1 + e^{i k_1}+e^{i k_2 })$.
If we divide $\gamma_{\bk } = i (\gamma_{c} (\bk )+ i \gamma_{s} (\bk
))$ into its even and odd components
\begin{eqnarray}\label{}
\gamma_{c} (\bk )&=& K (1 + \cos k_{1}+ \cos k_{2}), \cr
\gamma_{s} (\bk )&=& K (\sin k_{1}+ \sin k_{2}),
\end{eqnarray}
then  $g(i\omega_n)$ can be rewritten  as
	\begin{equation}
		g(z)=\frac{1}{N_c}\sum_{{\bf k}\in {\bf
		BZ}}\frac{z -(\gamma_c({\bf
		k})\tau_2+\gamma_s(\bf{k})\tau_1)}{z^2-\vert
		\gamma_{\bf k} \vert^{2}}.
		\label{GreensFunction}
	\end{equation} 
The odd component  $\gamma_s(\bf k)$ vanishes under momentum
summation so that 
	\begin{align}
		\begin{split}
		1-\hat{V}g(z)&=1-\frac{\xout{2} V}{N_c}\sum_{{\bf k} \in  \mathbf{BZ}}\frac{z \tau_2-\gamma_c({\bf k})}{z^2-\vert \gamma_{\bf k}\vert^2}
			\\&= 1-V(\tau_2g_0(z)-\mathbb{I}_2g_2(z))
		\end{split}
		\label{scattering_matrix}
	\end{align}
	where
\begin{eqnarray}\label{GreensFunctionDecompositionX}
	g_{0} (z)&\equiv& \frac{1}{N_c}\sum_{{\bf k} \in  \mathbf{BZ}}\frac{z}{z^2-\vert \gamma_{\bf k}\vert^2 },\cr
	g_2(z)&\equiv &\frac{1}{N_c}\sum_{{\bf k} \in \mathbf{BZ}}\frac{\gamma_c({\bf k})}{z^2-\vert \gamma_{\bf k} \vert^2}.
	\label{GreensFunctionDecompositionX}
\end{eqnarray}

Carrying out the trace in the free energy we then obtain 
\begin{eqnarray}
\label{free_energy_change_2}	
\Delta F&=&-\frac{T}{2}\tr\bigl[ \ln(1-\hat{V}G_{0})\bigr ]\\
			&=& -\frac{T}{2} \sum_{i\omega_n}\ln
			\left[(1+V g_2(i\omega_n))^2-(V
			g_0(i\omega_n))^2\right].\nonumber
	\end{eqnarray}
The Matsubara summation can then be carried out 
as an anti-clockwise contour integral around the imaginary axis 
weighted by Fermi function, $f (z)=[e^{\beta z}+1]^{-1}$. Deforming 
the contour to run clockwise around the real axis we obtain
\begin{equation}\label{deltaF}
\Delta F= \int_{-\infty }^{\infty }\frac{d\omega}{2\pi}\left(\frac{1}{2}-f
(\omega) \right) \delta_{v} (\omega),
\end{equation}
where
 \begin{widetext}
\begin{equation}\label{phase_shift}
\delta_v(\omega)= {\rm Im}\ \ln\biggl[(1+2 Kg_2(z))^2-(2 K g_0(z))^2\biggr]_{z=\omega-i \delta}
\end{equation}
\begin{figure*}[htp]
	\includegraphics[width=\columnwidth]{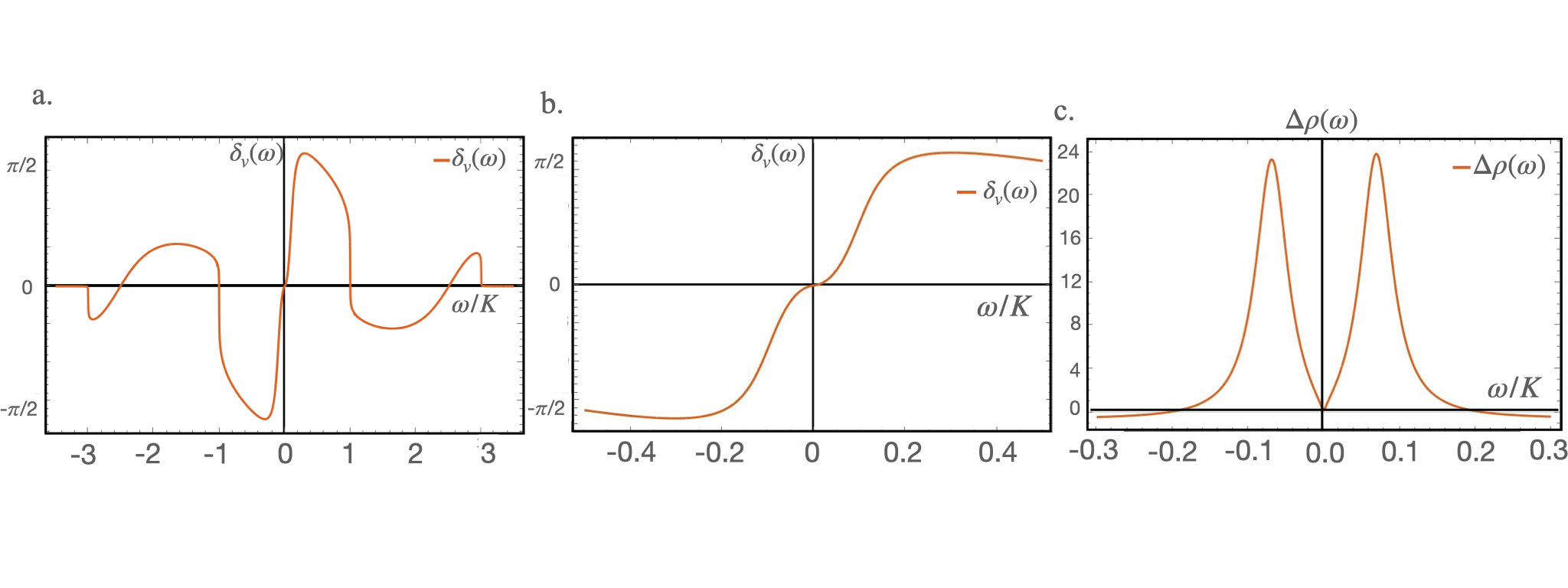}
	\caption{ (a)The scattering phase shift $\delta_v(\omega)$
		associated with the creation of two adjacent visons, as a
		function of frequency $\omega$ in units of $K$.(b)Scattering
		phase shift  $\delta_v(\omega)$ on an expanded scale, showing
		inflection point at origin. (c)Resonance in the scattering
		density of states  around $\epsilon_0=\pm 0.07K$ in the density
		of state change $\Delta \rho(\omega)$ due to the bond flip
		potential, as a function of frequency $\omega$ in units of
		$K$. This resonance will become sharp in the gapped topological
		state, when the resonance  drops below the gap
			edge, forming a sharp, in-gap excitation of the vison pair. }
	\label{PhaseShift}
\end{figure*}

is identified as the scattering phase shift, and $ g_0(z)_{z=\omega-i\delta}$ and $ g_2(z)_{z=\omega-i\delta}$, are the components of local Green's function $g(z)\vert_{z=\omega-i\delta}$(see Fig[\ref{GreensFunctionFig}]). Note that 
$\delta_{v} (\omega)=-\delta_{v} (-\omega)$ is an antisymmetric function of frequency.
At zero temperature 
the vison gap is then
\begin{eqnarray}\label{ztemp}	        
\Delta_v&=& K\int_{0}^{\infty} \frac{dx}{2\pi}\ 
{\rm Im}\ln  \biggl[(1+2 g_{2} (z))^{2}- (2 g_{0} (z))^{2} \biggr]_{z=x-i\delta }
	\end{eqnarray}
where we have rescaled the frequency in units of $K$, setting
$z=\omega/K$. In the reciprocal basis
	\begin{align}
		\begin{split}
			g_0(z)=\int_{0}^{2\pi}\frac{d k_1}{2\pi} \int_{0}^{2\pi}\frac{d k_2}{2\pi}\frac{z}{z^2-\vert \gamma_{\bf k}\vert^2},
			\\ g_2(z)=\int_{0}^{2\pi}\frac{d k_1}{2\pi} \int_{0}^{2\pi}\frac{d k_2}{2\pi}\frac{\gamma_c({\bf k})}{z^2-\vert \gamma_{\bf k}\vert^2},
		\end{split}
		\label{G0G2}
	\end{align}
where we have set 
$K=1$ in $\gamma (\bk )$, i.e $\gamma_{\bk }= 1+
e^{ik_{1}}+ e^{ik_{2}}$ and $\gamma_{c}= \cos (k_{1})+ \cos (k_{2})$.
	The interior integral over $k_2$ can be carried out as a
	complex contour
	integral over $w = e^{ik_{2}}$ around the unit circle, (Appendix \ref{AppendixA}), giving 
	\begin{align}
			g_0 (z)=
\int_{0}^{2\pi}\frac{d
			k}{2\pi}
			\frac{z}
{
(z^2-(3+2c))
\sqrt{1-\frac{8(c+1)}
{(z^2-(3+2c))^2}}
},
		\label{intG0}
	\end{align}
	\begin{align}
	g_2({z})=
\int_{0}^{2\pi}\frac{d k}{2\pi} \frac{2c+1}{({z}^2-(3+2c))\sqrt{1-\frac{8(c+1)}{({z}^2-(3+2c))^2}}},
		\label{intG2}
	\end{align}
	where $c\equiv \cos(k)$.	These integrals were evaluated numerically, to obtain the 
phase shift $\delta_v(\omega)$ (Fig[\ref{PhaseShift}]).  The phase shift
was interpolated over over a discrete set of $N$ points and the integral 
\eqref{ztemp} was carried out numerically on the interpolated phase
shift. By extrapolating the 
limit $1/N\rightarrow 0$, we find the vison gap energy to be
$\Delta_v=0.1311656(3) K=0.263313(6) J$ for the isotropic case 
$J_{z}=J_{y}=J_{z}=J$. 

This analytically-based calculation improves on the earlier
result obtained via numerical diagonalization of finite size systems
\cite{KSL1} i.e. $\Delta_{v}\approx 0.267 J$.  Its main virtue
however, is that the method can be easily generalized, and we gain
insights from the calculated scattering phase shifts. 
\end{widetext}

\begin{figure}[htp]
		\includegraphics[width=\columnwidth]{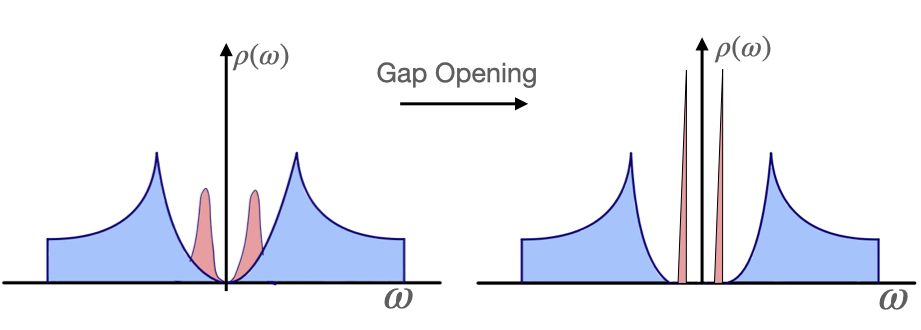}
		\caption{Schematic illustration of the resonance in
		the density of states for the gapless Kitaev spin
		liquid. The resonance becomes sharp when
		a gap opens in the bulk density of states, forming a
		fermionic excitation of the vison pair. }
	\label{SharpResonanceSchematic}
	\end{figure}

From the calculated phase shift, we can calculate the change
in density of states (DOS) (see Appendix \ref{AppendixB})
\begin{equation}
	\Delta\rho(\omega)=\frac{1}{\pi}\frac{d\delta_v }{d\omega}
	\label{DOSchange}
\end{equation}
(Fig. \ref{PhaseShift}
c.) associated with a bond flip, 
which is seen to contain a resonance centered around $\epsilon_0\approx  \pm
0.07 K$. 
This resonance can be examined in detail
by expanding $g_0(z)$ and
$g_2(z)$ for small $z$:

	\begin{align}
	\begin{split}
		g_0(\omega- i\delta)&=\frac{\omega}{\sqrt{3}\pi}\ln\left(\frac{3}{\vert \omega \vert}\right)+i\frac{\vert\omega \vert}{\sqrt{3}}
		\\ g_2(\omega- i\delta)&=-\frac{2}{3}-\frac{\omega^2}{3\sqrt{3}\pi}\left[\ln\left(\frac{3}{\vert\omega \vert}\right)+i\pi \sgn\ {\omega}\right]
	\end{split}
\end{align}
Which can be used to evaluate scattering phase shift
$\delta_v(\omega)$ (\ref{phase_shift}), and the resonant DOS change
$\Delta \rho (\omega)$ (\ref{DOSchange}) analytically.  The position
of the resonance is determined by the integration over the entire band
but its width is determined by the density of states at low
energies. Since the DOS vanishes inside the spectral gap, the
resonance will become sharp in the gapped topological phase when its
center lies beneath the gap edge (see Fig[\ref{SharpResonanceSchematic}]). The sharp peak in the gapped state signifies the binding of Majorana fermions to the visons formed by $\mathbb{Z}_2$ bond flip at origin. 
	\section{Discussion}
	In this work we have presented an analytical method for
determination of the vison gap by treating the flipping of the
$\mathbb{Z}_2$ gauge field as a scattering potential for the Majorana
fermions. In this way, we have been able to analytically extend the numerical
	treatment by Kitaev for the isotropic model on honeycomb
	lattice \cite{KSL1} to obtain an analytic result for the vison
	gap energy $\Delta_v$.
	
A key part of our approach is the calculation of the Majorana phase
shift for scattering off the bond-flipped configuration.  
One of the interesting observations is that the scattering contains a
Majorana resonance, located at an energy  $\epsilon_0 \approx  \pm 0.07 K $. Since this
resonance is formed from scattering throughout 
the entire Brillouin zone, its location is 
robust. Thus in those cases where the excitation spectrum acquires a
gap, eg through time-reversal symmetry breaking \cite{KSL1,haldane_model_1988}, this
resonance transforms into a sharp in-gap  excitation. 
\begin{figure}[htp]
	\includegraphics[width=\columnwidth]{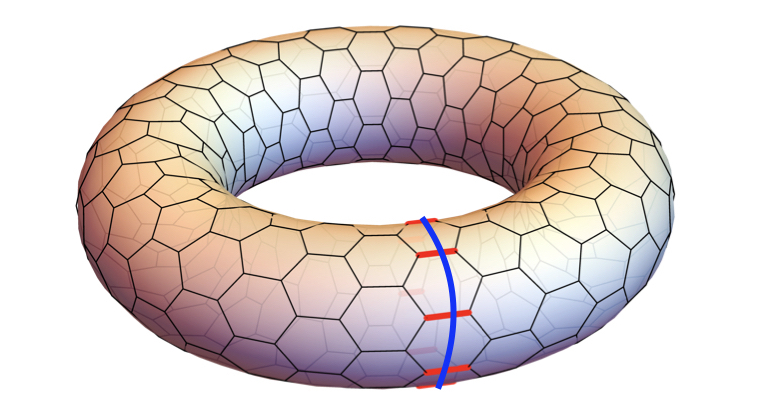}
	\caption{Hexagonal lattice of Kitaev spin liquid is embedded
	on a torus by application of periodic boundary condition. An
	anyons forms within the torus is formed flipping the bonds
	along a non-contractable loop that encircles the torus. }
	\label{TorusAnyon}
\end{figure}

{ While it is possible to extend our method to
analytically calculate the energy associated with the injection of an
anyon into the torus, by flipping
$x-x$ bonds along ${\bf{a_1}}$ direction(see Fig[\ref{TorusAnyon}]), a much simpler derivation of
the anyon energy in the KSL can be
made by taking two copies of the KSL,
\begin{align}
	\begin{split}
	H_{KSL,1}=2\sum_{<ij>} J_{\alpha_{ij}}\hat{u}_{ij}(i c_{1i} c_{1j})
	\\ H_{KSL,2}=2\sum_{<ij>} J_{\alpha_{ij}}\hat{u}_{ij}(i c_{2i} c_{2j})
	\end{split}
\end{align}
forming a complex fermion Hamiltonian
$H_c=(H_{KSL,1}+H_{KSL,2}=2\sum_{<ij>} J_{\alpha_{ij}}\hat{u}_{ij}
(i c^{\dagger}_{i} c_{j}+{\rm H.c})$, where $c_i\equiv
(c_{1i}+ic_{2i})/\sqrt{2}$ is a complex fermion. The ground state
excitation energy of $H_c$ corresponding to the gauge configuration
where one reverses of $x-x$ bonds along ${\bf{a_1}}$ direction is
twice the anyon energy for $H_{KSL}$.
For $H_c$ the line of reverse bonds around
the torus can then be absorbed by a unitary transformation that
redistributes the odd boundary condition into an effective vector
potential that shifts all the momenta 
${\bf k}=(k_1,k_2)\rightarrow(k_1+\frac{\pi}{L},k_2)$,
equivalent to introducing a half magnetic flux with 
vector potential $A_{x}= \frac{\pi}{L}$. 
Treating the response to the vector potential in an analogous fashion
to a superconductor, 
the putative the energy cost of an anyon would be
 \begin{equation}\label{}
\Delta E=\int d^2 x \frac{\rho_s}{4}
 A_{x}^2= \rho_{s}\frac{\pi^2}{4}  ,
\end{equation}
 where $\rho_s$ is the superfluid stiffness associated with the
 ground-state, $A= \pi/L$ is the vector  potential and the factor of
 $4$ derives from halving the energy of the complex fermion system. 
However, since the complex fermion Hamiltonian $H_{c}$ preserves the
global $U (1)$ symmetry, 
its superfluid stiffness $\rho_s$ vanishes 
so it costs no energy to create anyons in the gapless state. From this
line of reasoning, we can see that the ground state of the Kitaev spin
liquid has a four-fold degeneracy and is topologically ordered.

The method can also be applied to calculate the energy of visons separated by finite distance by treating the bond flips between the two visons as a diffraction problem, however due to complex nature of the problem, it is left for future investigation.
	
Finally, we note that our method also admits various
	generalizations.  For example, it can be extended 
to anisotropic couplings i.e. $J_x\neq J_y\neq
	J_z$ as well as to higher dimensions, such as the
	three-dimensional  hyperoctagonal lattice.
Moreover, our method can be applied to study the impact of
	spinor order formation as a consequence of hybridization
	between conduction electrons and Majorana spinons in the
	CPT model for a Kondo lattice coupled to a Yao-Lee spin
	liquid	\cite{tsvelik_order_2022,coleman_solvable_2022}. This allows
	us to study the stability of Yao-Lee spin liquid against
	spinor order formation, the subject of a forthcoming article by the authors. 
	
	\begin{acknowledgments}
	This work was supported by Office of Basic Energy Sciences, Material
 Sciences and Engineering Division, U.S. Department of Energy (DOE)
 under Contracts No. DE-SC0012704 (AMT) and DE-FG02-99ER45790 (AP and PC ).
{All authors contributed equally to this work.} 
	\end{acknowledgments}
\appendix
	\section{Analytic Calculation of Green's Function in Honeycomb Lattice}
	\label{AppendixA}
Here we show how to simplify the integrals 
	\begin{align}
		\begin{split}
			{g}_0(z)=
\int_{0}^{2\pi}\frac{d k_1}{2\pi} \int_{0}^{2\pi}\frac{d k_2}{2\pi}\frac{z}{z^2-\vert \gamma_{\bf k}\vert^2},
			\\ {g}_2(z)=\int_{0}^{2\pi}\frac{d k_1}{2\pi} \int_{0}^{2\pi}\frac{d k_2}{2\pi}\frac{\gamma_c({\bf k})}{z^2-\vert \gamma_{\bf k}\vert^2},
		\end{split}
		\label{G0G2}
	\end{align}
\noindent where $\gamma_c ({\bk })=1+ \cos(k_1)+\cos(k_2)$, using a contour
integral.  We begin by noting that the integrals over $k_{1}$ and
$k_{2}$ can be carried out in either order, allowing us to pull the
cosines in $\gamma_{c} (k)$ out of the inner integral, so that 
\begin{eqnarray}\label{}
g_{0} (z)&=& \int_{0}^{2\pi}\frac{d k_1}{2\pi} z I_{0} (z,k_{1}),\cr
g_{2} (z)&=& \int_{0}^{2\pi}\frac{d k_1}{2\pi} (1+2 \cos k_{1}) I_{0} (z,k_{1}),
\end{eqnarray}
where
\begin{equation}\label{}
I_{0} (z,k) = \int_{0}^{2\pi} \frac{dk_{2}}{2\pi} \frac{1}
{z^2-\vert \gamma_{\bf k}\vert^2}.
\end{equation}
Writing  $s=e^{i k_1}$ and $w = e^{i k_2}$, we can rewrite
$I_{0}$ as a counter-clockwise integral around the unit circle $|w|=1$, 
	\begin{align}
I_{0} (z,k)\equiv I_{0} (z,s)= \ointctrclockwise\limits_{\vert w \vert=1}\frac{d
w}{2\pi i w} \frac{1}{z^{2}-|\gamma (s,w)|^{2}}.
	\label{contourgreens}
	\end{align}
Rewriting the denominator
as a quadratic function of $w$,
		\begin{align}
			\begin{split}
				\small	z^2-\vert \gamma_(s,w)\vert^2&=z^2-(1+s+w)(1+\frac{1}{s}+\frac{1}{w})
				\\&=-\frac{(1+s)}{s w}(w^{2}+ wb + s),
			\end{split}
		\end{align}
where
\begin{equation}\label{}
b = \frac{1+3s + s^{2}-s z^{2}}{(1+s)}.
\end{equation}
We can thus write the integral in the form 
\begin{equation}\label{}
I_{0} (z,s) = -\frac{s}{1+s}\ointctrclockwise \frac{dw}{2\pi i} \frac{1}{(w-w_{+}) (w-w_{-})}
\end{equation}
where 
\begin{equation}\label{roots}
w_{\pm} = -\frac{b}{2}\pm \sqrt{\left(\frac{b}{2} \right)^{2}-s}
\end{equation}
are the poles of the integrand. 

Now since $w_{+}w_{-}=s=e^{ik_{1}}$,
it follows that $|w_{+}w_{-}|=1$, so that only one of these poles 
lies inside the contour. (In general, this may depend on the way we
treat the branch
cuts inside the square root of \eqref{roots}. However, we 
don't actually need to know which pole it
is, as this we will fix the sign and the branch-cuts 
in the final expression by demanding that the asymptotic behavior of
$I_{0}\sim 1/z^{2}$ is analytic at large $z$.) Lets assume that the pole
closest to the origin is at $w=w_{-}$, then
we obtain 
\begin{equation}\label{}
I_{0} (z,s)= \frac{s}{1+s}\frac{1}{w_{+}-w_{-}} = \frac{s}{1+s}\frac{1}{\sqrt{b^{2}-4s}}.
\end{equation}
\begin{widetext}
Now expanding the denominator, we have 
\begin{eqnarray}\label{}
(1+s)\sqrt{b^{2}-4s} &=& \sqrt{(1+3s+s^{2}-sz^{2})^{2}-4s
(1+s)^{2}}\cr
&=& s\sqrt{(3+ 2 \cos k_{1}-z^{2})^{2}- 8 (\cos k +1)}\cr
&=& s (z^{2}- (3+ 2 \cos k_{1})) 
\sqrt{1- 
\frac{ 8 (\cos k +1)}
{(z^{2}- ( 3+ 2 \cos k_{1}))^{2}}},
\end{eqnarray}
where we have factorized the final expression, 
to guarantee that at 
large $z$, $I_{0} (z,s)\sim
1/z^2$ is analytic. 
Combining the above results, 
gives us the following expressions for $g_{0} (z)$ and $g_{2} (z)$
	\begin{align}
		\begin{split}
			{g}_0({z})=
\int_{0}^{2\pi}\frac{d k}{2\pi} \frac{{z}}{({z}^2-(3+2c))\sqrt{1-\frac{8(c+1)}{({z}^2-(3+2c))^2}}}
			\\	{g}_2(z)=\int_{0}^{2\pi}\frac{d k}{2\pi} \frac{2c+1}{({z}^2-(3+2c))\sqrt{1-\frac{8(c+1)}{({z}^2-(3+2c))^2}}}
		\end{split}
		\label{intG0G2}
	\end{align}
	Where  $c\equiv \cos(k)$, which are the expressions given in
	\eqref{intG0} and \eqref{intG2}.
\end{widetext}
		\section{Density of scattering states around a vison
}\label{AppendixB}

In this appendix we discuss the interpretation of the scattering phase
shift in a Majorana scattering problem.  For conventional particles,
the scattering phase shift is defined in terms of the S matrix, $S =
e^{2i \delta }$ associated with the scattering of a partial wave
state: what is the appropriate generalization to Majorana
excitations? We can answer this question by considering the scattering
Hamiltonian.  In our problem, the scattering Hamiltonian is given by
\eqref{SHam} which we rewrite in momentum space as 
\begin{equation}\label{SHam2}
_{KSL+2v}=
\frac{1}{2}\sum_{{\bf k}\in {\bf BZ}} {\psi }^{\dagger}_{{\bf
k}}(\vec{\gamma}_{{\bf k}}\cdot \vec{\tau}) \psi _{{\bf k}}
+\frac{1}{2}\sum_{\bk ,\bk '\in BZ}{\psi }\dg_{\bk' } (V\tau_2 ){\psi }_{\bk }.
	\end{equation}
Here the $\psi_{\bk }$, which are the Fourier transform of the 
real-space Majorana fermions, are complex ``Dirac'' fermions;
their underlying Majorana character is enforced by the
condition $\psi_{-\bk \alpha }= \psi\dg_{{\color{red}\bk} \alpha }$,
so that holes formed in one half of the Brillouin zone are equivalent
to particles in the other half. 
This guarantees that the density of states $\rho (E)=\rho (-E)$ is
particle-hole symmetric. 
The factor
of $1/2$ in the Hamiltonian avoids overcounting. 

An independent set of one-particle excitations can be formed in two ways.
\begin{enumerate}
\item By using all
of momentum space, but restricting the excitations exclusively to
positive energy, particle excitations,  i.e one-particle eigenstates
are 
\begin{equation}\label{}
\vert \bk \rangle  = \sum_{\alpha } u_{\bk \alpha }\psi \dg_{\bk \alpha }\vert GS\rangle,
\qquad (
\bk \in BZ, \ \alpha \in A,B)
\end{equation}
where $u_{\bk \alpha }\psi \dg_{\bk \alpha }$ creates a positive
energy eigenstate at momentum $\bk $.
This allows us to discuss the phase shift of scattered, conventional
fermions.  The density of states is then $N_{I} (E)= \rho (E)
$ and the free energy is given by an integration over positive energy excitations
\begin{equation}\label{symmresult}
F =-T \int_{0}^{\infty }dE \ln (1+e^{-\beta  E})\rho (E),
\end{equation}

\item Alternatively, restricting $\bk$ to
one half of the Brillouin zone while considering both particles and
holes formed within this half subspace. In this case, the density of
states is $N_{II} (E)= \frac{1}{2}\rho (E)$ and the free energy is
written 
\begin{eqnarray}\label{}
F &=&-\frac{T}{2} 
\int_{-\infty }^{\infty } dE\ln (1+e^{-\beta  E})\rho
(E)\cr
&=& -\frac{T}{2}\int_{-\infty }^{\infty }dE \ln (2 \cosh (\beta  E/2) )\rho
(E)
\end{eqnarray}
where $\rho (E)=\rho (-E)$. 

\end{enumerate}
Method 1. is more appropriate for discussing the scattering, while
2. is more aligned with the Greens function approach we have
adopted.

To calculate the change in density of states due to the scattering, we
note that the change in Free energy calculated in \eqref{deltaF}
\begin{equation}\label{deltaF}
\Delta F= \int_{-\infty }^{\infty }\frac{d\omega}{2\pi}\left(\frac{1}{2}-f
(\omega) \right) \delta_{v} (\omega),
\end{equation}
can be integrated by parts to obtain 
\begin{equation}\label{}
\Delta F = -\frac{T}{2}\int_{0}^{\infty } dE \ln \left[2 \cosh
\frac{\beta  E}{2} \right]
\left(\frac{1}{\pi}\frac{\partial \delta (E)}{\partial E} \right).
\end{equation}
Comparing this with \eqref{symmresult}, we see that the change in
density of states due to scattering is
\begin{equation}\label{}
\delta \rho (E) =  \frac{1}{\pi}\frac{\partial \delta
(E)}{\partial E}
\end{equation}

We note that this result can also be obtained by obsrving that 
the  scattering  shifts the energy
levels $E_{\lambda}$ of the continuum downwards by an amount equal to 
\begin{equation}\label{}
E'_{\lambda} =E_{\lambda} - \Delta  \frac{\delta (E_{\lambda})}{\pi}
\end{equation}
where $\Delta$ is the energy spacing of the continuum. 
The energy spacing is then modified by the scattering, to
\begin{eqnarray}\label{}
\Delta '&=&E'_{\lambda+1}-E'_{\lambda}=\Delta  - \frac{\Delta 
}{\pi} [\delta (E_{\lambda}+\Delta )-\delta (E_{\lambda})]\cr
&=& \Delta \left[1  - \frac{\Delta }{\pi}\frac{\partial\delta
}{\partial E} \right].
\end{eqnarray}
If the original density of states is $\frac{1}{\Delta}= \rho (E)$, the
modified density of states is then 
\begin{eqnarray}\label{}
\rho  (E) + \delta \rho (E) &=& \frac{1}{\Delta '} = \frac{1}{\Delta }\left[1  - \frac{\Delta }{\pi}\frac{\partial\delta
}{\partial E} \right]^{-1}\cr &=& \rho (E)+
 \frac{1}{\pi}\frac{\partial \delta
(E)}{\partial E}
\end{eqnarray}
so that the change in the density of states is given by 
\begin{equation}\label{}
\Delta \rho (E) = \frac{1}{\pi }\frac{\partial \delta (E)}{\partial E}.
\end{equation}

%

\end{document}